\documentstyle[12pt,epsf]{article}

\newcommand{\be}{\begin{equation}}
\newcommand{\ee}{\end{equation}}
\newcommand{\bea}{\begin{eqnarray}}
\newcommand{\eea}{\end{eqnarray}}

\def\double{\baselineskip 18pt \lineskip 10pt}

\begin{document}

\vspace{1in}

\begin{center}
\Large
{\bf What is the homogeneity of our universe telling us?}

\vspace{0.4in}

\normalsize

\large{Mark Trodden\footnote{trodden@erebus.cwru.edu} and Tanmay 
Vachaspati\footnote{tanmay@theory4.cwru.edu}}

\normalsize
\vspace{.3in}

{\em Particle Astrophysics Theory Group \\ Department of Physics \\
Case Western Reserve University \\ 10900 Euclid Avenue \\ Cleveland, OH 
44106-7079, USA}

\end{center}

\vspace{.3in}

\begin{abstract}
The universe we observe is homogeneous on super-horizon scales,
leading to the ``cosmic homogeneity problem''. Inflation alleviates
this problem but cannot solve it within the realm of conservative
extrapolations of classical physics. 
A probabilistic solution of the problem is possible but is
subject to interpretational difficulties.
A genuine deterministic solution 
of the homogeneity problem requires radical departures from known 
physics.
\end{abstract}

\vspace{.3in}

\begin{center}
{\em Awarded Honorable Mention in the 1999 Gravity Research Foundation Essay
Competition.}
\end{center}

\vfill
\begin{flushright} 
CWRU-P20-99
\end{flushright}
\eject

\double 

It is well known that the standard cosmological model has a homogeneity 
problem: why is the temperature of the cosmic microwave background 
the same to a high degree of accuracy in regions that have never been in 
causal contact?
%mark -according to the traditional theory of cosmic evolution?
This observation is one of the primary motivations for the concept of 
{\it inflation} 
%mark
\cite{Guth,Linde}, which has been the foundation
of a revolution in cosmology over the last two decades.
The rapid expansion of cosmic inflation can 
stretch an initially small, smooth spatial region 
to a size much larger than the observable universe today, providing
a hope of explaining the present day large-scale 
homogeneity of the universe.

While the observed homogeneity is a compelling 
reason to study inflation in detail,
the mechanism can only truly be said to solve the homogeneity problem if the
initial smooth region from which inflation begins is causally correlated, so
that its own homogeneity is achievable by physical processes. 
Otherwise, we are left
with the problem of understanding the homogeneity of the initial inflating
patch; another, albeit less severe, homogeneity problem.

It is then a striking result that, under certain conservative
assumptions, if the universe is not born inflating, large-scale 
homogeneity is required for inflation to begin \cite{VacTro98}. 
Hence classical inflationary models which employ 
conservative extrapolations of known physics, can only be considered 
to alleviate, but not solve, the homogeneity problem. 
Furthermore, the extensions to known physics that are required to
solve the homogeneity problem are quite novel and provide
hints that may be used to construct new physical theories.

The main constraint on inflationary models comes from the 
requirement that gravitational forces not be ``too repulsive''. 
This constraint can be embodied in the 
Raychauhuri equation governing the divergences of light rays 
({\it i.e.} null geodesics) which says that, if 
$\theta = \nabla_a N^a$ denotes the divergence of a congruence 
of null geodesics whose tangent vectors are $N^a$, then
\begin{equation}
{{d\theta} \over {d\tau}} + {1\over 2} \theta^2 =
-\sigma_{ab}\sigma^{ab} + \omega_{ab}\omega^{ab}
-R_{ab}N^a N^b \ ,
\label{raychaudhari}
\end{equation}
where $\tau$ is the affine parameter along the null geodesic, 
$\sigma_{ab}$ is the shear tensor, $\omega_{ab}$ the twist tensor
and $R_{ab}$ the Ricci tensor. For a specially
chosen congruence of null rays - one that is hypersurface 
orthogonal - it can be shown \cite{Wald} that,
\begin{equation}
{{d\theta} \over {d\tau}} \leq -R_{ab}N^a N^b = - 8\pi T_{ab}N^a N^b \ ,
\label{rayc}
\end{equation}
where $T_{ab}$ is the energy-momentum tensor, and in the last equality we 
have used Einstein's equations in natural units.

The weak energy condition concerns the energy-momentum
tensor of the matter.  This condition is satisfied by all known matter at
the classical level, and it seems reasonable to assume that
it should be satisfied generally. A straightforward consequence is
\begin{equation}
T_{ab}N^a N^b \geq 0 \ ,
\label{wec}
\end{equation}
which for a perfect fluid amounts to requiring a positive
energy density, $\rho \geq 0$, and a pressure that is bounded
from below by minus the energy density: $ p \geq - \rho$.
The Raychaudhuri equation, in conjunction with the weak energy condition, 
then leads to
\begin{equation}
{{d\theta} \over {d\tau}} \leq 0 \ .
\label{usedcondition}
\end{equation}
This equation is a form of the physical statement that
the gravitational forces between reasonable matter should not
be too repulsive.  Were negative energy densities or arbitrarily 
large negative pressures allowed in the theory, the statement 
would not be true.

We want to understand the implications of the constraint 
(\ref{usedcondition}) for inflationary models. 
Consider a universe in which, due to causal processes, 
a small patch is undergoing inflation but
is immersed in a spacetime which itself may be expanding
but not inflating. Then there are null rays that originate in
the background spacetime and enter the inflating region.
We can calculate $\theta$ in both the background region 
and the inflating region. In fact, if the expansion of both
regions is given by a scale factor ($a(t)$) as in the standard
cosmology, radially incoming null rays have a divergence 
given by
\begin{equation}
\theta = {2 \over {a(t)}} \left ( H - {1\over x} \right ) \ ,
\label{thetafrw}
\end{equation}
where, as usual, $H = {\dot a}/a$ and
$x$ is the physical radial distance of the ray at time $t$.
Eq. (\ref{usedcondition}) implies that $\theta$ cannot
be negative in the background region and positive in the
inflating region which, when used in conjunction with
(\ref{thetafrw}), gives:
\begin{equation}
H_{\rm inf}^{-1} \geq H_{\rm FRW}^{-1} (t_i) \ ,
\label{hdshfrw}
\end{equation}
where $t_i$ is the time that inflation started and the
subscripts refer to the inflating (inf) and non-inflating
(FRW) spacetimes.

It seems reasonable to assume that the conditions leading to
inflation must be satisfied over a region larger than 
$H_{\rm inf}^{-1}$. Then, from eq. (\ref{hdshfrw}), the patch size 
that can inflate to form our observable universe
has to be larger than the background Hubble scale,
$H_{\rm FRW}^{-1} (t_i)$. Note that $H^{-1}$ is large compared
to typical length scales over which particle interactions can 
homogenize the 
universe. Hence, large-scale homogeneity has to
be an initial condition for cosmic inflation to proceed, and therefore
such inflation does not solve the homogeneity problem.
This is the striking result alluded to earlier.

Nevertheless, the universe does exhibit large-scale homogeneity, the only
proposed explanation for which is inflation.
It is therefore worthwhile considering what it
takes to genuinely solve the homogeneity problem in the context of
inflationary models. The derivation
of the above result does not hold if at least one of the following
statements is true:
\begin{itemize}
\item There exist violations of the classical Einstein equations, 
say due to quantum effects.
\item The weak energy condition is violated in the early universe.
\item The universe has non-trivial topology. 
\item The universe is born directly into an inflating universe,
that is, there is no pre-inflationary epoch, such as might 
occur in quantum cosmology.
\item Singularities other than the big bang are present.
\end{itemize}

Probably the most conservative approach is to consider quantum
effects in the early universe. We think this is conservative
because we know that quantum mechanics correctly describes the
world we live in, whereas the other possible options require
conditions that are not seen today. However, a quantum mechanical
explanation of the homogeneity is necessarily a probabilistic
solution to the problem, and is subject to differing interpretations
since we observe only one universe. When considering a quantum mechanical 
origin of the homogeneity, one must consider both the possibility of directly
producing the observed universe, and the possibility of 
generating the appropriate inflationary initial conditions from which it 
could evolve.  The principle behind adopting inflation as a paradigm is that 
it greatly enhances the probability for the creation of the
universe that we see. However, this does leave open the issue of the 
probability of producing inflationary initial conditions themselves.
An analogy might help clarify
this situation. Suppose there are a hundred coins laid out on a
table and we find that all of them have their heads facing up.
Should we then say that the coins were thrown at random and
we are simply seeing a highly unlikely chance event? Or should
we say that, at a later time, someone carefully arranged
the coins with their
heads up? Inflation is analogous to the latter case here. However, it can
only be viable if we understand the probability of a process that can ``turn 
all the coins face up''. In the cosmological context there are anthropic 
considerations that confuse the interpretation yet further -
it may be that we can only see a coin if its head 
is facing up. Such questions are extremely difficult to answer
and, at present, it is fair to say that no convincing answer
is known. The same difficulties (together with other technical
ones) arise when one attempts to explain the creation of our universe 
by quantum cosmology \cite{Vil82,HarHaw83}.

%Even if we ignore the difficulties of a probabilistic interpretation
%of obtaining cosmic inflation via quantum processes, it is not
%yet certain if quantum effects can give rise to appropriate
%departures from the Einstein equations or suitably large 
%violations of the weak energy condition that will lead to inflation.

Faced with the difficulties of a probabilistic interpretation
of obtaining cosmic inflation via quantum processes, we may consider 
less conservative directions.
If new kinds of matter are present that couple in novel ways to
the metric, they can either modify Einstein's equations such that the
last equality in (\ref{rayc}) does not hold, or else they can provide
violations of the weak energy condition (``extremely repulsive matter'').
% In this case, it
%might be possible for inflation to solve the homogeneity problem.
It is interesting to note that non-minimally coupled scalar fields are
a specific example of matter that can evade our constraint. 
Such fields arise naturally in supergravity and string theory, a possible
quantum theory of gravity. 
It may be that the observed homogeneity
%General Relativity 
is steering us to consider these fields as promising inflaton candidates. 
%Alternatively,
%quantum mechanics may be responsible for violations of the weak energy
%condition, although such effects are constrained by the Ford-Roman 
%inequalities \cite{For}.
In addition, if space has non-trivial topology \cite{glenn}, 
we may also recover an inflationary solution to the
homogeneity problem. In this case, however, the length scale
associated with cosmic topology should be comparable to the inflating
horizon size.

Another escape from the result is possible if we include singularities
other than the big bang
in the spacetime. Here, however, the singularity must border the
inflating patch of the universe. So, even
if one did produce an inflationary patch, there would be no way of
predicting events in this patch without first understanding the
nature and influence of the singularity \cite{FarGutGuv,BorTroVac98}.

To summarize, the observed homogeneity problem cannot have an
inflationary solution within conventional extrapolations of 
classical physics. If one 
wishes to find a solution, novel departures from classical physics must 
be considered. The quantum solution is beset with interpretational 
difficulties. Other classical solutions rely on violations of the weak 
energy condition or modifications of Einstein's equations 
that would, in effect, provide a strongly repulsive
gravitational event in the history of the universe.

\end{document}